\documentclass[sort&compress
    ,final            
  ]
  {aipproc}
\usepackage{sidecap,graphicx,boxedminipage,epsfig,wrapfig,floatflt,shadow}
\layoutstyle{8x11double}

\begin{document}

\title{Student understanding of Symmetry and Gauss' law}


\author{Chandralekha Singh}{
  address={Department of Physics, University of Pittsburgh, Pittsburgh, PA, 15213}
}

\begin{abstract}
Helping students learn why Gauss' law can or cannot be easily applied to determine the strength of the electric field 
at various points for a particular charge distribution, and then helping them learn to determine the shape of the Gaussian
surfaces if sufficient symmetry exists can develop their reasoning and problem solving skills. 
We investigate the difficulties that students in calculus-based introductory physics
courses have with the concepts of symmetry, electric field and electric flux that are pivotal to Gauss' law of electricity. 
Determination of the electric field using Gauss' law requires discerning the symmetry of a particular
charge distribution and being able to predict the direction of the electric field everywhere if a high symmetry exists. 
It requires a good grasp of how to add the electric field vectors using the principle of superposition, and
the concepts of area vector and electric flux.
We administered free response and multiple-choice questions and conducted interviews with individual students using 
a think-aloud protocol to elucidate the difficulties students have with the concepts of symmetry, electric field and electric flux.
Here, we discuss student responses to some questions on a multiple-choice test administered to them. The test can be used as a teaching and assessment tool.
\end{abstract}

\maketitle

\vspace{-0.35in}
\section{Introduction}
\vspace{-0.1in}

A major goal of most calculus-based introductory physics courses is to help students develop the thinking skills of a physicist~\cite{alan}.
Physicists comfortably exploit symmetries to turn complicated problems into simpler ones. 
Gauss' law of electricity 
is an important topic in the second semester of most calculus-based introductory physics courses.
Learning to reason whether Gauss's law can be exploited in a particular situation to calculate the electric field without complicated integrals
can provide an excellent context for helping students develop a good grasp of symmetry considerations. 
Unfortunately, students often memorize a collection of formulas for the magnitude of the electric field 
obtained using Gauss's law for spherical, cylindrical, and planar geometries without paying attention to symmetry 
considerations involved in obtaining those results or even understanding the difference between the electric field and
electric flux. Therefore, a majority of students have difficulty identifying situations 
where Gauss' law may be useful and overgeneralize results obtained for a highly symmetric charge distribution to situations 
where they are not applicable. 
Courses often do not sufficiently emphasize symmetry considerations or the chain of reasoning required to 
determine if Gauss' law is useful for calculating the electric field.

We administered free response and multiple-choice questions and conducted interviews with individual students using a
think-aloud protocol~\cite{chi} to understand the difficulties students have with the concepts of symmetry, electric field and electric flux.
Then, we developed a conceptual multiple-choice test that addresses these issues and administered it to 168 students in three different courses.
The tests and interviews assess the extent to which students have become proficient in exploiting symmetry and in making conceptual predictions
about the magnitude and direction of the electric field for a given charge distribution using Coulomb's or Gauss' laws. 
They also assess whether students can distinguish between electric
field and electric flux, identify situations in which Gauss' law can easily be used to calculate the electric field strength
from the information about the electric flux and the shapes of the Gaussian surfaces that would be appropriate in those cases.

\vspace{-.28in}
\section{Discussion}
\vspace{-.1in}

Investigation of student difficulties related to a particular physics concept is 
important for designing instructional strategies to reduce them~\cite{mcdermott}.
Our investigations of student difficulties with discerning symmetry and applying Gauss' law was performed
using two methods: (1) design and administration of free-response and multiple-choice questions to elicit
difficulties in a particular context, and (2) in-depth audio-taped
interviews with individual students using a think-aloud protocol~\cite{chi} while they solved those problems. The
major advantage of written tests is that they can be administered to large
student populations.  When administered in conjunction with
in-depth interviews with a subset of students, written tests can be very helpful in
pinpointing student difficulties. 
While a student can always calculate the electric field using Coulomb's law 
and their knowledge of the principle of superposition without regard to symmetry, 
determining electric field using Gauss' law requires an explicit focus on the symmetry of the charge distribution.
Although there are only three types of symmetry (spherical, cylindrical, and planar) for which Gauss' 
law can easily be exploited to determine the electric field at various points from the information about the electric flux, 
students need help in identifying when these symmetries are present. 
The principle of superposition is also a pre-requisite for employing Gauss' law successfully to determine the electric field
and helps determine if sufficient charge symmetry exists in a particular situation. 

In addition to the consideration of symmetry, 
the area vector and the electric flux are new concepts that are introduced 
in the context of Gauss' law.
Students must be able to distinguish the electric flux from the electric field, a task that turns out to be extremely difficult. 
Students must learn that the electric flux, a measure of the total number of electric field lines passing through an {\it area}, 
is a scalar (which can have both positive and negative signs depending upon the relative directions of the electric field and area vector) 
while the electric field at a point is a vector. 
They must understand that the electric flux and the electric field even have different dimensions. 

Students must also understand that Gauss' law only applies to closed surfaces and that for any closed surface, information about the net charge
enclosed is sufficient to determine the net electric flux through it. On the other hand, determination of the electric field at various
points due to a charge distribution is not simple because it depends on the exact manner in which charges are 
distributed and may depend upon both the charges inside and outside the closed surface. For example, the electric field 
at various points on a Gaussian surface may be non-zero and vary from point to point 
even though the net electric flux through it is zero if the net charge enclosed is zero. 

During the design of the multiple-choice test on symmetry and Gauss' law, we paid particular attention to the important issues of reliability 
and validity~\cite{nitko}. Reliability refers to the relative degree of consistency between testing if the test procedures were repeated for 
an individual or group. Validity refers to the appropriateness of test score interpretation.
The test design began with the development of a test blueprint which provided a comprehensive framework for
planning decisions about the desired test attributes. 
The rest of the test design proceeded in a manner similar to that described for the Energy and Momentum Conceptual Survey~\cite{energy},
and will be described in detail elsewhere.

The final version of the test contains 25 items. It was administered to a total of 168 students in three different introductory calculus-based 
physics courses at Pitt. One of these was an honors course, and the students in the other two courses had used some preliminary tutorials on 
symmetry and Gauss' law.
The average score on the test is approximately $56\%$. The reliability coefficient  $\alpha$~\cite{nitko} (a measure of the internal 
consistency of the test) is $0.77$ which is considered good from the test design standards. The point biserial
discimination is related to the ability of a question to discriminate between students who overall did well on the test vs. those whose overall 
performance on the test was poor. This discrimination index for 16 questions on the test was more than $0.4$. Only 4 questions had discrimination
index less than $0.3$ (only one question with less than 0.2) which is also very good from the test design standards~\cite{nitko}.
Below, we discuss some of the questions on the test:

\vspace*{.1in}
{\bf {$\bullet$} \underline{Setup for the next two questions}}\\
You perform two experiments (see figure) in which you distribute charge $+6Q$ differently on an isolated hollow insulating sphere: \\
Exp(I): You put identical charges, $+Q$ each, on the spherical surface in six localized blobs (you can consider them point charges)
such that the adjacent blobs are equidistant from each other.\\
Exp(II): You distribute charge $+6Q$ uniformly on the surface of the sphere.
\begin{figure}
\includegraphics[height=.1\textheight]{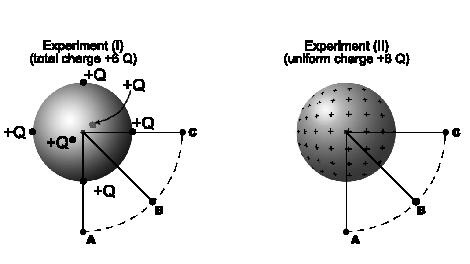}
\caption{Figure for questions (1) and (2)}
\end{figure}

\begin{enumerate}
\item
In both Exp (I) and (II), points A, B, and C are \underline{equidistant} from the center and lie in the same equatorial plane of the sphere.
In Exp (I), points A and C are straight out from two of the charges and point B is in between points A and C as shown.\\

Which \underline{one} of the following statements is true about the electric field magnitudes at points A, B, and C due to the $+6Q$ charge?\\

(A) In each experiment, the field magnitude at points A, B, and C is the same, but the magnitudes differ in the two experiments.\\
(B) In each experiment, the field magnitude at points A, B, and C is the same, and the magnitudes are equal in the two experiments.\\
(C) In experiment (I), the field magnitude is the same at points A, B, and C, but not in experiment (II).\\
(D) In experiment (II), the field magnitude is the same at points A, B, and C, but not in experiment (I).\\
(E) None of the above.\\

\item
Which \underline{one} of the following statements is true about the electric field magnitude at any point \underline{inside} the sphere
due to the $+6Q$ charge (see figure above)?\\

(A) It is zero everywhere inside the sphere in both experiments.\\
(B) It is non-zero everywhere inside the sphere in both experiments.\\
(C) In experiment (I), it has the same non-zero magnitude everywhere, but it is zero everywhere inside the sphere in experiment (II).\\
(D) In experiment (I), it has a magnitude that varies from point to point, but it is zero everywhere inside the sphere in experiment (II).\\
(E) None of the above.\\


{\bf {$\bullet$} \underline{Setup for the next two questions}}\\
A cubic Gaussian surface with \underline{1 meter} on a side is oriented with two horizontal and four vertical faces, as shown.
The cube is in a uniform electric field of \underline{20 N/C} which is directed vertically upward.
Point A is on the top surface and point B on a side surface of the Gaussian cubic surface.\\
\begin{figure}
\includegraphics[height=.1\textheight]{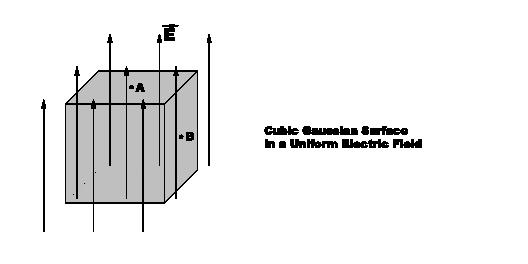}
\caption{Figure for questions (3) and (4)}
\end{figure}
\item
Which \underline{one} of the following statements is true about the electric field at points A and B?\\
(A) The field is zero at both points A and B.\\
(B) The field is zero at point A but not at point B.\\
(C) The field is zero at point B but not at point A.\\
(D) The field is non-zero at both points A and B and its direction is the same at the two points.\\
(E) The field is non-zero at both points A and B but its direction is different at the two points.\\

\item
List all of the following statements that are true about the electric flux.\\
\hspace*{0.45in} (I) The net flux through the surface of the\\
\hspace*{0.74in} whole cube is zero.\\
\hspace*{0.45in} (II) The magnitude of the flux through the 
\hspace*{0.74in} top face of the cube is $20 N m^2 /C$.\\
\hspace*{0.45in} (III) The magnitude of the net flux 
\hspace*{0.76in} through the whole cube is $20 N m^2 /C$.\\
(A) (I) only\\
(B) (II) only\\
(C) (III) only\\
(D) (I) and (II) only\\
(E) (II) and (III) only\\

{\bf {$\bullet$} \underline{Setup for the next two questions}}\\

Shown below are four imaginary surfaces coaxial with an isolated infinitely long line of charge (with uniform linear charge density $\lambda$):\\
\hspace*{0.45in} (I) a closed cylinder of length $L$.\\
\hspace*{0.45in} (II) a sphere of diameter $L$.\\
\hspace*{0.45in} (III) a closed cubic box with side $L$.\\
\hspace*{0.45in} (IV) a two dimensional square sheet with 
\hspace*{0.74in} side $L$. The line of charge is perpen
\hspace*{0.74in}  -dicular to the plane of the sheet.\\
\begin{figure}
\includegraphics[height=.1\textheight]{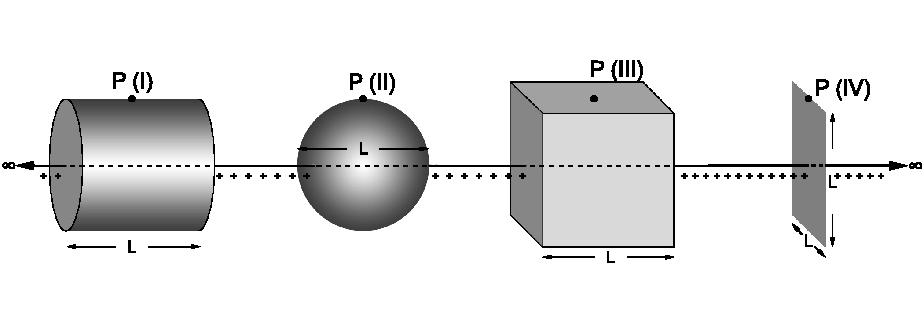}
\caption{Figure for questions (5) and (6)}
\end{figure}
\item
List all of the \underline{above surfaces} through which the net electric flux is $\phi=\lambda L/\epsilon_0$.\\

(A) (I) only\\
(B) (I) and (II) only\\
(C) (I) and (III) only\\
(D) (I), (II), and (III) only\\
(E) (I), (II), (III), and (IV)\\

\item
List all of the \underline{above surfaces} that can be used as Gaussian surfaces to easily find the \underline{electric field} magnitude
(due to the infinite line of charge) at a point P shown on the surface using Gauss' law:\\

(A) (I) only\\
(B) (I) and (II) only\\
(C) (I) and (III) only\\
(D) (I), (II), and (III) only\\
(E) (I), (II), (III), and (IV)\\

\end{enumerate}

\begin{table}[h]
\centering
\begin{tabular}[t]{|c|c|c|c|c|c|c|}
\hline
$Item \#$ & A & B& C&D&E&$\%$ Correct \\[0.5 ex]
\hline \hline
1& 39&15&4&\it{96}&14&57\\
\hline
2&60&7&\it{84}&6&11&50\\
\hline
3&4&4&56&\it{97}&7&58\\
\hline
4&36&18&8&\it{99}&7&59\\
\hline
5&48&8&22&\it{64}&26&38\\
\hline
6&\it{71}&23&32&27&15&42\\
\hline
\end{tabular}
\caption{The number of students (out of a total of 168) who selected choices (A)-(E) on items (1)-(6) and the percent of students who chose correct answer. The correct response for each question has been italicized.} \label{table1}
\end{table}

\vspace*{-0.2in}
The most common difficulty with question (1) was assuming that the magnitude of the electric field in experiment (I) will also be the same at all of the three points shown.
In interviews and free response questions, some students explained this by claiming that since the six point charges and the three points are symmetrically 
situated, the field magnitude must work out to be the same at all the three points shown while others explained their response by claiming that for a point 
outside, the six point charges on the sphere can be thought to be point charges at the center of the sphere.
The most common difficulty with question (2) was assuming that the electric field inside the sphere in experiment (I) is also zero everywhere.
In interviews and free response questions, some students explained their response by claiming that the electric field inside a sphere (or inside
an object of any shape with charges on its surface) is always zero. When probed further, some students mentioned that this had to do with electrical
shielding which guaranteed that the electric field will be zero inside. Even when interviewers drew students' attention 
to the fact that the sphere was made of insulating material, students adhered to the idea that the field inside must be zero everywhere.

The most common difficulty with question (3) was assuming that the electric field is zero at point B on the side surface of the cube although 
the problem statement explicitly mentioned that the cube is in a uniform electric field of $20N/C$.
In interviews and free response questions, some students explicitly claimed that the area vector of the side surface is perpendicular to the
direction of the electric field lines. Therefore, the electric field must be zero at point B. This kind of confusion between the electric field
at a point and the contribution to the electric flux from a certain area was found in other questions as well.
The most common difficulty with question (4) was that students did not realize that both statements (I) and (II) are correct. The table above shows that many students
only chose option (I) or (II) but not both.

Questions (5) and (6) were extremely difficult for students. Students were not comfortable with the statement of Gauss' law that relates
the net flux through a closed surface to the net charge enclosed. 
They also could not differentiate between electric flux through a closed surface and
electric field at a point on the surface. For example, in question (5), many students chose (I) (or (I) and (III)) and claimed in interviews and free response questions that only those surfaces can be used to determine the net electric flux through them
because the other surfaces did not have the correct
symmetry.
Some students were also unsure about the distinction between open and closed surfaces and the fact that Gauss' law only applies to closed surfaces. 
For example, in interviews and free response questions, many students claimed that the net flux should be the same through all surfaces (including
the square sheet)
using Gauss' law. In Question (6), students had to choose the Gaussian surfaces that would help them determine the electric field at
point P easily due to the infinite line of charge. All of the alternative choices were selected with an almost even frequency.  Students
were often unsure about the symmetry concepts relevant for making appropriate decisions and those who chose option (C) were often
quite confident that the magnitude of the electric field due to the infinite line must be the same at every point on the cube as well.

\vspace{-0.28in}
\section{Work in Progress}
\vspace{-0.15in}

Gauss' law provides an excellent opportunity to help students learn symmetry ideas and to develop their reasoning and problem solving skills.
We are currently implementing and refining several tutorials
that we have developed to help students learn to 
discern the symmetry of a charge distribution, determine whether Gauss' law can easily be applied to calculate the 
electric field strength at various points due to a given charge distribution, and predict the shape of the corresponding Gaussian surface if a high
symmetry exists. The tutorials also help students learn to distinguish between the electric field and the electric flux, understand that 
the electric flux is not a vector while the electric field is, learn the superposition principle and determine if the electric field 
at two points is the same in magnitude and/or direction due to a given charge distribution. 
The tutorials in this series also help students calculate the electric field at various points due to highly symmetric charge distributions and
revisit superposition principle after learning Gauss' law, for example, to determine the net electric field due to two spheres of charge.
All tutorials have separate pre/post-tests that students take before and after working on the tutorials. The preliminary results are encouraging.

\bibliographystyle{aipproc}   

{}


\IfFileExists{\jobname.bbl}{}
 {\typeout{}
  \typeout{******************************************}
  \typeout{** Please run "bibtex \jobname" to obtain}
  \typeout{** the bibliography and then re-run LaTeX}
  \typeout{** twice to fix the references!}
  \typeout{******************************************}
  \typeout{}
 }

\end{document}